\documentclass[11pt,a4paper]{article}
\pdfoutput=1
\usepackage{jheppub}
\usepackage{epsfig,simplewick,xcolor,amsmath}
\usepackage{fancybox}
\usepackage{graphicx, caption}

\newcommand{\be}{\begin{equation}}
\newcommand{\bea}{\begin{eqnarray}}
\newcommand{\ee}{\end{equation}}
\newcommand{\eea}{\end{eqnarray}}

\begin{document}
\title{A Two-Phase Perspective on Deep Learning Dynamics}
\author[a,b]{Robert de Mello Koch\footnote{\href{mailto:robert@zjhu.edu.cn}{robert@zjhu.edu.cn}}}
\affiliation[a]{School of Science, Huzhou University, Huzhou 313000, China}
\affiliation[b]{Mandelstam Institute for Theoretical Physics, School of Physics, University of the Witwatersrand, Private Bag 3, Wits 2050, South Africa}
\author[a]{and Animik Ghosh\footnote{Corresponding author; \href{mailto:animikghosh@gmail.com}{animikghosh@gmail.com}}}
\date{April 2025}
\abstract{We propose that learning in deep neural networks proceeds in two phases: a rapid curve-fitting phase followed by a slower compression or coarse-graining phase. This view is supported by the shared temporal structure of three phenomena -- grokking, double descent, and the information bottleneck -- all of which exhibit a delayed onset of generalization well after training error reaches zero. We empirically show that the associated time scales align in two rather different settings. Mutual information between hidden layers and input data emerges as a natural progress measure, complementing circuit-based metrics such as local complexity and the linear mapping number. We argue that the second phase is not actively optimized by standard training algorithms and may be unnecessarily prolonged. Drawing on an analogy with the renormalization group, we suggest that this compression phase reflects a principled form of forgetting, critical for generalization.}
\maketitle

\section{Introduction}

Grokking is the phenomenon where a neural network, after memorizing its training data and showing poor generalization, suddenly and dramatically begins to generalize well -- typically long after achieving zero training error \cite{Grok1,Grok2,Grok3}. Originally observed in synthetic tasks, recent studies suggest that grokking is surprisingly widespread and can arise in a range of practical settings \cite{AlwaysGrok}.

A compelling reason to study grokking is that it challenges classical learning theory. It exposes a mismatch between fitting and understanding: networks can memorize perfectly yet take much longer to internalize the correct algorithm or structure, as seen in tasks like modular arithmetic. This delayed transition reveals a richer set of dynamics underlying generalization than standard bias-variance tradeoffs capture \cite{AlwaysGrok,Grok4,Grok5,Grok6}.

Grokking also shows that training dynamics remain active well after the loss plateaus. Internal representations may continue evolving in subtle ways that ultimately enable generalization. Understanding these dynamics can shed light on hidden mechanisms governing learning, help us better navigate the optimization landscape, and inform practical decisions like early stopping, regularization, and learning rate schedules.

From a research perspective, grokking provides a tractable, interpretable setting -- what one might call the ``fruit fly" of generalization. It allows us to study deep learning dynamics in a clean environment where signals like compression, abstraction, and memorization can be precisely tracked. In this paper, we argue that grokking is a manifestation of a deeper structure: that learning proceeds as a two-stage process -- first through curve fitting, and later through a phase resembling coarse graining or ``forgetting''. We make this argument by identifying a common structure between three phenomena: grokking, double descent, and the information bottleneck principle.

To set the stage, we briefly review the relevant aspects of double descent and the information bottleneck. Double descent refines the classical view of generalization by showing that test error, after initially rising near the interpolation threshold, can decrease again as model complexity increases. This non-monotonic behavior has been observed across a wide range of models, including neural networks, kernel methods, and even simple linear regression. The information bottleneck, meanwhile, offers a principled framework for understanding how a model balances compression and predictive power. It posits that the optimal internal representation of data is one that retains as much relevant information about the target as possible, while discarding irrelevant input features.

We propose that these frameworks are not merely analogous but reflect a common underlying structure, inducing the same fundamental phases in the learning process. Specifically, we demonstrate that three critical time scales—the point at which grokking achieves perfect training accuracy, the onset of the second descent in double descent, and the transition to compression in the information bottleneck—are empirically aligned. In Section \ref{Time scale}, we define these time scales, explain their significance, and outline how they can be measured in practice. Section \ref{Numerics} presents the empirical evidence supporting this alignment. In Section \ref{2-phase learning}, we motivate the hypothesis that learning unfolds in two distinct stages: an initial curve-fitting phase, followed by a second phase resembling coarse graining or ``forgetting". This picture is consistent with detailed numerical studies of grokking. A key outcome of our work is the proposal that the mutual information between inputs and outputs of a hidden layer serves as a natural progress measure for deep network training. Finally, in Section \ref{Discussion}, we discuss the implications of our findings and suggest directions for future work.

\section{Time Scales and their relevance} \label{Time scale}

Grokking offers a tractable and interpretable setting for studying generalization in deep neural networks. The insights it provides directly motivate the central hypothesis we present in Section \ref{2-phase learning}. To build toward that, we review key supporting results in this section. Section \ref{Grokking} revisits the main observations about grokking, which already hint that learning unfolds in two distinct phases. To identify the first of these phases, we turn to the phenomenon of double descent -- an update to the classical bias-variance trade-off -- which we review in Section \ref{DD}. The second phase, characterized by a form of compression, is naturally described by the information bottleneck framework, reviewed in Section \ref{IB}.

\subsection{Grokking} \label{Grokking}

Grokking refers to the phenomenon in which a deep neural network (DNN) begins to generalize well only after a prolonged period of training, long after achieving near-zero training error. By now, several detailed dynamical studies have explored grokking, offering an appealing and intuitive understanding of some of its underlying mechanisms~\cite{Grok1,explgrok1,Grok4}. A key concept we will build on is that of a \emph{circuit}~\cite{explgrok2}. To motivate this idea, we begin by recalling that a DNN applies a sequence of nonlinear transformations to an input vector $\vec{x}$, propagating it through $L$ layers. This defines a function $f_{\theta}(\vec{x})$ of the form
\begin{equation}
f_{\theta}(\vec{x}) \equiv a\left(W^{(L)}a\Bigg(\cdots a\Big(W^{(2)} a(W^{(1)}\vec{x}+b^{(1)})+b^{(2)}\Big)\cdots +b^{(L-1)}\Bigg)+b^{(L)}\right)
\end{equation}
where $\theta$ denotes the complete set of network parameters. Each layer $l$ contributes a weight matrix $W^{(l)}_{ij}$ and a bias vector $b^{(l)}_i$ to $\theta$. The activation function $a(\cdot)$ is a nonlinearity applied element-wise to the output of each neuron. In our case, we use the ReLU activation function,
\begin{equation}
a(t) = \max(0, t),
\end{equation}
which outputs zero whenever $t < 0$. This implies that such neurons are effectively inactive and do not contribute to the network’s output. As a result, for any given input, the computation performed by the network involves only a subset of neurons and connections.

\begin{figure*}[h]
\includegraphics[width=1.0\linewidth]{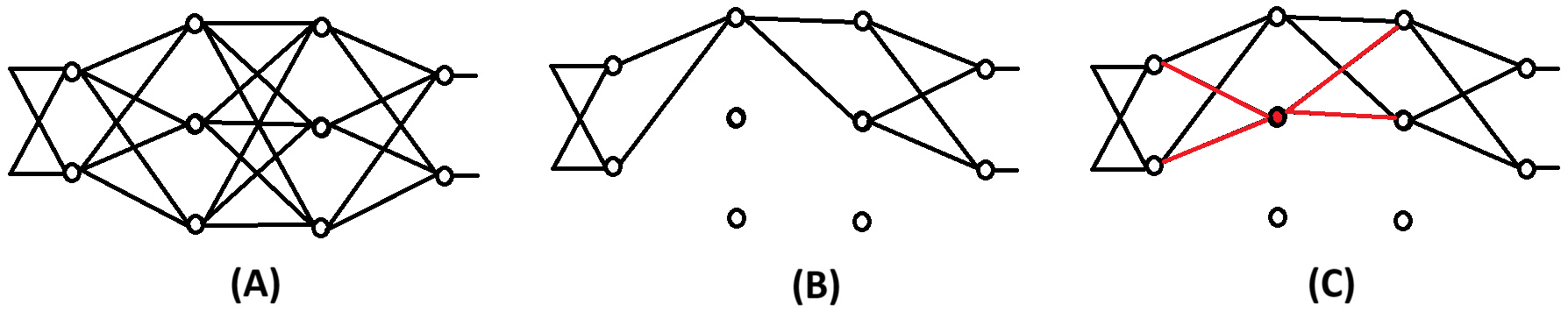}
\caption{\textbf{Circuits in a ReLU MLP} \textbf{(A)} Schematic of a multi-layer perceptron (MLP) with an input and output layer, and four hidden layers. \textbf{(B)} A representative activation pattern in which only the top neuron in the second hidden layer and the top two neurons in the third hidden layer are active. The resulting \emph{circuit} includes only the activated neurons and the weights connected between them. \textbf{(C)} Transition to a neighboring linear region corresponds to the addition or removal of a single neuron from the circuit. In the example shown, the second neuron in the second hidden layer becomes newly activated.}
\label{fig:Circuits}
\end{figure*}

This idea can be formalized using the notion of a \emph{circuit}, as illustrated in Figure~\ref{fig:Circuits}. A circuit consists of the subset of neurons that are activated for a given input, along with the weights that connect them. For that input, the circuit defines a linear mapping from the input to the output. Let $x^{(l)}_i$ denote the input to neuron $i$ in layer $l$. Because we are using the ReLU activation function, the boundary between active and inactive neurons is determined by the zero locus of the pre-activation:
\begin{equation}
\mathcal{H}_i^{(l)} = \left\{ x^{(l)}_i : \sum_j W^{(l)}_{ij} x^{(l)}_j + b^{(l)}_i = 0 \right\}.\label{precativationlocus}
\end{equation}
Each \( x^{(l)}_i \) can be regarded as a function of the network input \( \vec{x} \), so that for a fixed set of parameters \( \theta \), the input space is partitioned into regions according to the activation pattern of the neurons. Within each region, the network computes a distinct linear function. Since moving between neighboring regions corresponds to crossing one of the zero-locus boundaries \( \mathcal{H}_i^{(l)} \), adjacent circuits differ by the activation or deactivation of a single neuron.

\begin{figure*}[h]
\includegraphics[width=1.0\linewidth]{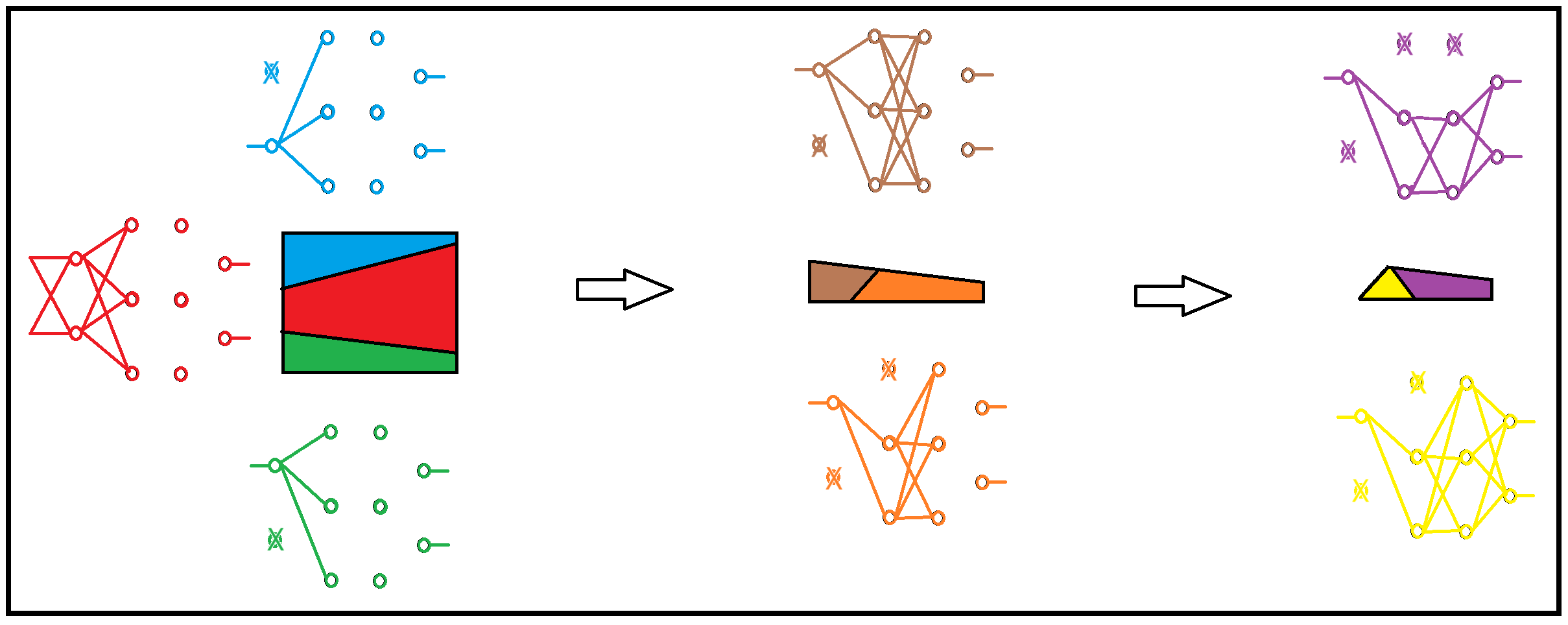}
\caption{\textbf{Partitioning input space:} The ReLU MLP of Figure \ref{fig:Circuits} has two input parameters. This input space is visualized as a square above. The first partitioning (left hand side) is performed using the activation locus of neurons in the first layer, the second partitioning (middle) using neurons in the second layer and the third (right hand side) using neurons in the third layer. The square is divided into a blue and a red region by the line indicating where the upper neuron becomes inactive and into a red and a green region where the lower neuron becomes inactive. The three regions blue, red and green, are associated with the circuits of the same color. Inactive neurons are marked with a $\times$. The green region is then considered. It is dissected into two regions, by the line for which the upper neuron in the second layer becomes inactive. There are similar partitions for the blue and red regions that we have not spelled out. Finally, the orange region is considered. It is dissected into two regions by the line on which the upper neuron in the third layer becomes inactive. There is one more partition, not illustrated, by the neurons in the final layer. The result is that the input space is partitioned into regions with a unique circuit associated to each region. Circuits in adjacent regions differ by a single neuron. }
\label{fig:partition}
\end{figure*}

It is therefore useful to view the input space as partitioned into regions, each associated with a distinct circuit. A simple illustration of such a partition is shown in Figure~\ref{fig:partition}. A \emph{progress measure}~\cite{ProgMeas} for DNN training is a scalar quantity that ideally evolves monotonically with training. One natural progress measure can be defined in terms of circuits~\cite{AlwaysGrok,Grok5}. By counting the number of distinct circuits within a fixed-volume neighborhood of the input space, we obtain a measure of the network’s \emph{local complexity} (LC). During training, the number of distinct circuits typically decreases, leading to a reduction in the integral of LC over the input space.

A closely related progress measure, known as the \emph{Linear Mapping Number} (LMN), was introduced in~\cite{LMN}. The construction begins with a \emph{linear connectivity matrix} $L_{ij}$, whose entries are based on the square of Pearson's correlation coefficient\footnote{The correlation coefficient between two random variables ranges from $-1$ to 1. An absolute value of 1 implies that a linear equation describes the relationship between the random variables perfectly. A value of 0 implies that there is no linear dependency between the variables.}, and whose rows and columns are indexed by samples in the input space. If two samples are strongly linearly related, the corresponding entry in $L_{ij}$ is close to 1. Diagonal entries are set to 1 by definition, while weak or nonlinear relationships result in values closer to 0.

The eigenvalues \( \lambda_i \) of \( L_{ij} \) are then used to define a normalized spectrum:
\begin{equation}
\tilde{\lambda}_i = \frac{|\lambda_i|}{\sum_j |\lambda_j|}.
\end{equation}
Using these normalized eigenvalues, the LMN is defined as:
\begin{equation}
\mathrm{LMN} = 2^{S_{\mathrm{NL}}}, \qquad S_{\mathrm{NL}} = -\sum_i \tilde{\lambda}_i \log_2 \tilde{\lambda}_i,
\end{equation}
where \( S_{\mathrm{NL}} \) can be interpreted as the entropy of the normalized spectrum. In simple examples~\cite{ProgMeas}, it can be shown that the LMN corresponds exactly to the number of distinct linear regions in the input space. As a result, the LMN is closely related to the integral of the LC over the input space.

Beyond offering a natural progress measure, circuits provide a remarkably compelling lens through which to view the dynamics of grokking \cite{AlwaysGrok}. By the time the network achieves zero training error, the complexity of the input space is largely concentrated near the training samples. At this point, one might reasonably assume that the network undergoes little further change. However, this intuition is misleading: both the LMN and the integrated LC continues to decrease even after the training loss has saturated. During this phase, the complexity of the linear regions migrates away from the training data samples, effectively smoothening the network’s mapping in the vicinity of those points.

Tracking the location and evolution of linear regions throughout training reveals rich microscopic details about the grokking process. Initially, the network forms a highly complex mapping in an effort to minimize training error. Over time, this evolves into a mapping that is significantly smoother around the training examples -- an essential feature for generalization. This behavior strongly suggests a two-phase learning process. The first phase, dominated by curve fitting, ends when the training error reaches zero. The second phase begins thereafter and continues until the network generalizes successfully, corresponding to a minimum in complexity. We will return to this interpretation in more detail in Section~\ref{2-phase learning}.

\subsection{Double Descent} \label{DD}

Double descent \cite{dd1,dd2,dd3} challenges the classical understanding of how model complexity affects generalization. As model complexity increases, test error is expected to follow a U-shaped curve: it decreases at first as the model becomes flexible enough to fit the data, then increases as the model begins to overfit capturing noise or irrelevant patterns in the training set. This view is rooted in the bias-variance trade-off, where simple models suffer from high bias (underfitting) and complex models from high variance (overfitting).

In the double descent picture, something surprising happens after the peak of the classical U-curve. As model complexity continues to increase beyond the point where the model can exactly fit (interpolate) the training data, the test error begins to decrease again. This second phase of improvement leads to a second ``descent". The point where the model first achieves perfect training accuracy is known as the interpolation threshold, and it's around this region that test error tends to be at its worst. Pushing the model into the overparameterized regime -- where it has far more parameters than training data -- often leads to better generalization performance.

The classic U-curve is indicative of a curve fitting phase, while the second descent indicates a compression or coarse graining phase, in which irrelevant details of the data are discarded by the network. The time scale controlling the exit from curve fitting to coarse graining is determined by the peak of the U-curve. The second phase can be taken to terminate when the second descent halts.

\subsection{Information Bottleneck} \label{IB}

The Information Bottleneck (IB) is a theoretical framework for understanding and visualizing how learning systems extract relevant structure from data~\cite{IB1}. It proposes that a good internal representation $T$ of an input $X$ should balance two competing objectives: retaining useful information about a target $Y$, while discarding irrelevant details from $X$. This trade-off is formalized by minimizing the IB objective:
\begin{equation}
\mathcal{L}_{\text{IB}} = I(T; X) - \beta I(T; Y),
\end{equation}
where $I(A; B)$ denotes the mutual information between random variables $A$ and $B$, and $\beta \geq 0$ is a trade-off parameter. In our setting, $X$ represents the training inputs, $Y$ the corresponding labels, and $T$ may refer to the output of any hidden layer in the network. Mutual information measures how much knowing one variable reduces uncertainty about another. Accordingly, $I(T; X)$ quantifies how much information the representation $T$ contains about the input $X$, while $I(T; Y)$ measures how much predictive information $T$ retains about the labels. The aim is to minimize $I(T; X)$ (compress irrelevant input details) while maximizing $I(T; Y)$ (preserve predictive structure), resulting in compact and effective internal representations.

Empirical evidence suggests that hidden layers in deep networks approximately implement this trade-off during training~\cite{IB2,IB3}. A helpful visualization of these dynamics is provided by the \emph{information plane}, where each hidden layer is represented as a point with coordinates $(I(T; X), I(T; Y))$. As training progresses, the network’s weights evolve, inducing motion of these points across the information plane.

Training typically unfolds in two distinct phases. In the early \emph{fitting phase}, both $I(T; X)$ and $I(T; Y)$ increase as the network learns to capture structure in the input that helps minimize training error. This phase is relatively fast and corresponds to the network memorizing details of the training data. Once the training error becomes small, the network enters the \emph{compression phase}, during which $I(T; X)$ gradually decreases while $I(T; Y)$ remains roughly constant. This signals that the network is discarding superfluous input information—effectively ``forgetting" non-predictive details—while retaining what is necessary for accurate output prediction.

This transition marks a shift from memorization to abstraction, suggesting that generalization in deep learning depends not only on fitting but also on simplifying internal structure. The dynamics traced on the information plane exhibit a clear two-phase trajectory: the transition point at which $I(T; X)$ begins to decrease defines the onset of compression, and the end of this phase is marked by the minimum of $I(T; X)$.

Having now identified and defined the time scales associated with all three phenomena—grokking, double descent, and the information bottleneck—we are in a position to compare them directly. The next section presents numerical experiments in settings where all three effects can be observed and quantitatively analyzed.

\section{Numerical Results} \label{Numerics}
In this section, we substantiate our claims in the previous section with numerical evidence. We first describe our experimental setup in Subsection \ref{Setup} followed by our results in Subsection \ref{Results}.
\subsection{Experimental Setup} \label{Setup}
Our model architecture involves two transformer layers with 4 attention heads followed by a feedforward network with Gaussian Error Linear Unit (GeLU) activation 
\begin{equation}
    \text{GELU}(x) \sim 0.5x \left[1+\tanh\left[\sqrt{\frac{2}{\pi} (x+0.044715 x^3)}\right]\right]
\end{equation}
In the theoretical discussion presented in Section~\ref{Grokking}, we made extensive use of the ReLU activation function. This choice was primarily motivated by conceptual clarity: ReLU introduces a well-defined switching point at $x=0$, where a neuron transitions between active and inactive states. The trade-off, however, is that the ReLU function has a discontinuous first derivative at this point. In contrast, the GeLU activation function smooths out the transition, providing a continuous derivative that is more favorable for numerical optimization. The cost of this smoothness is the absence of a natural, sharply defined active/inactive boundary, which in turn leads to less sharply defined boundaries between linear regions. Our analysis does not rely on precise definitions of linear region boundaries, and the circuit framework remains a valid and useful conceptual tool regardless of the activation function used. Supporting this, we note that the Linear Mapping Number (LMN) can be evaluated just as easily for networks with ReLU or GeLU activations. We train our model on algorithmic datasets: division modulo 96 and the multiplication table of the permutation group $S_5$ in particular. We have trained the neural network parameters with AdamW optimizer (with learning rate = $10^{-3}$ and weight decay = 0.0) with cross-entropy loss for $10^6$ steps. We have used embedding dimension = 128 (i.e. each head acts on a 32 dimensional subspace of the embedding space) and hidden dimension = 512. We used a 40-60 train-test split for the modular division case and a 50-50 train-test split for the $S_5$ group.

\begin{figure}[h] 
    \centering
    \begin{minipage}[b]{0.4\textwidth}
        \centering
        \includegraphics[width=\linewidth]{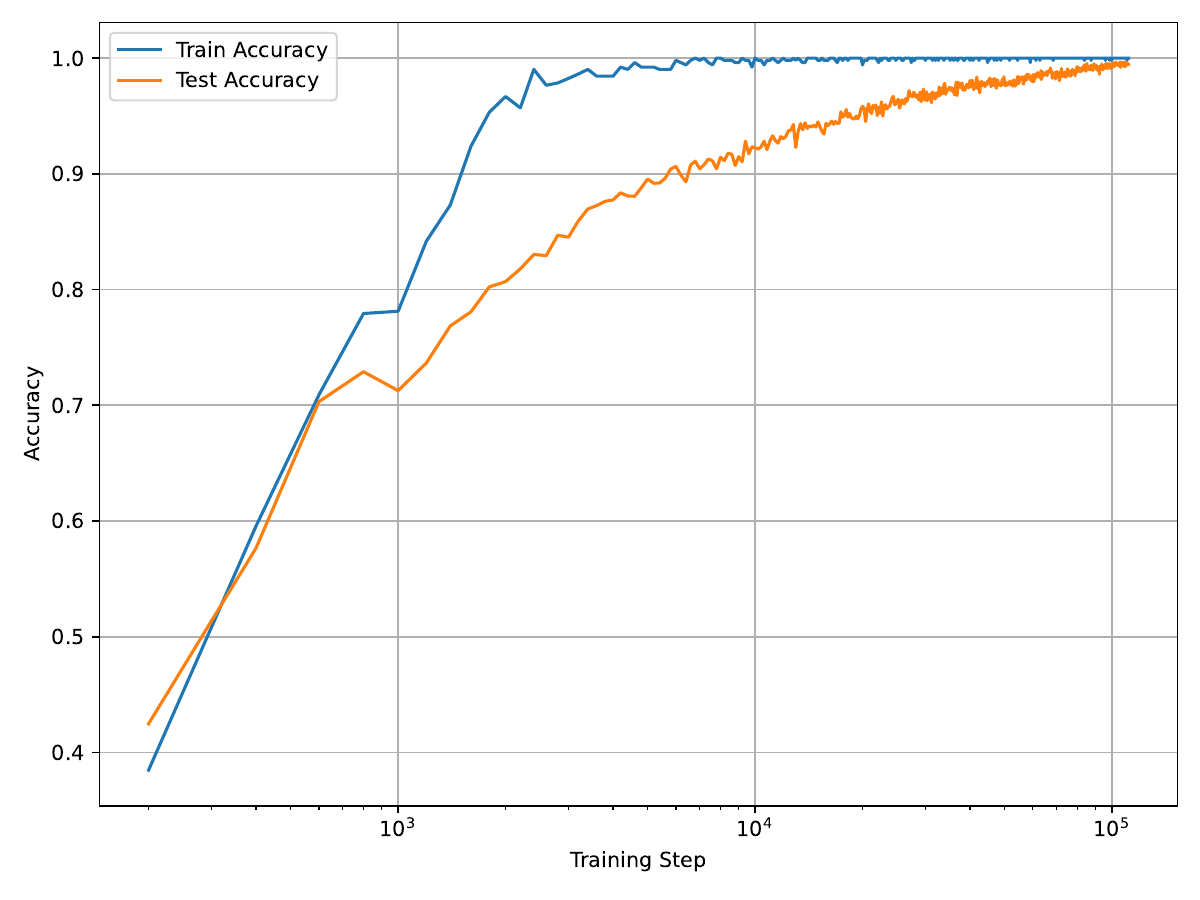}
        \caption*{\textbf{(a)} } 
    \end{minipage}
    \hfill
    \begin{minipage}[b]{0.4\textwidth}
        \centering
        \includegraphics[width=\linewidth]{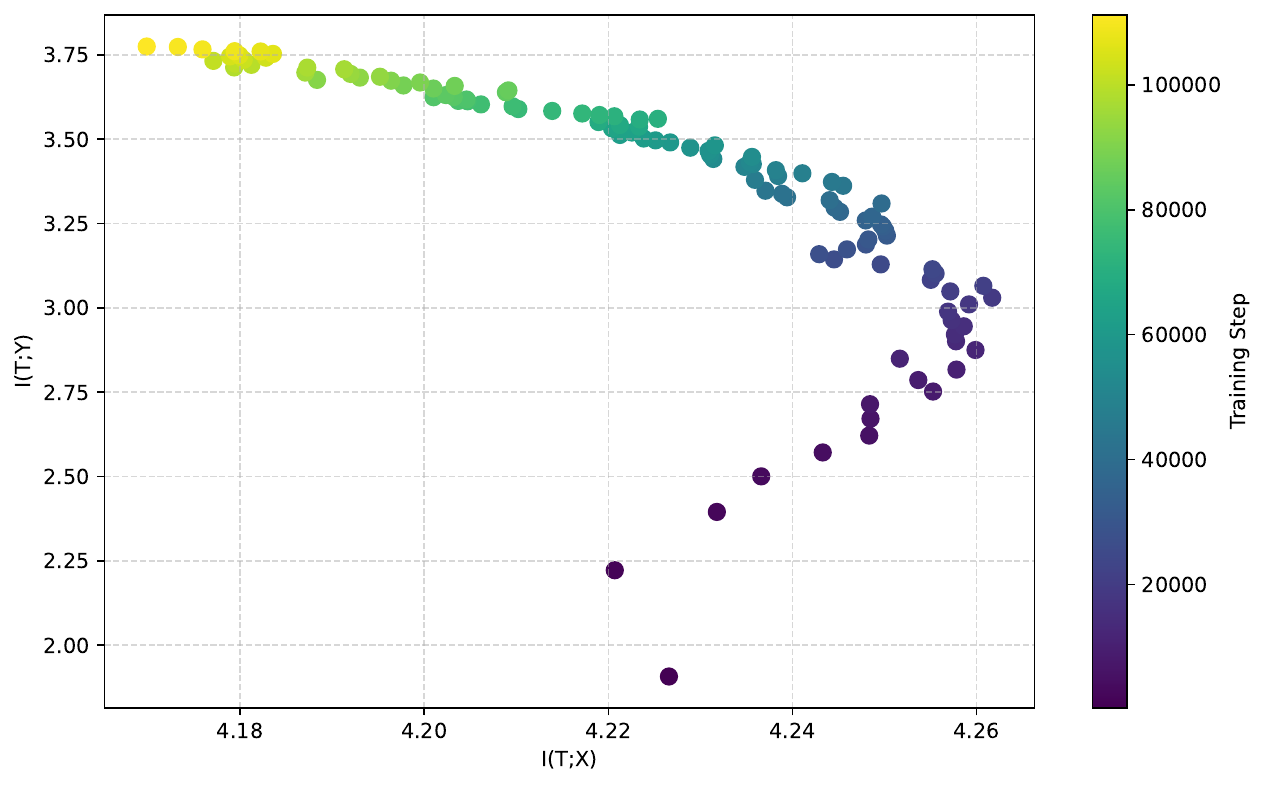}
        \caption*{\textbf{(b)} }
    \end{minipage}
    \hfill
    \begin{minipage}[b]{0.4\textwidth}
        \centering
        \includegraphics[width=\linewidth]{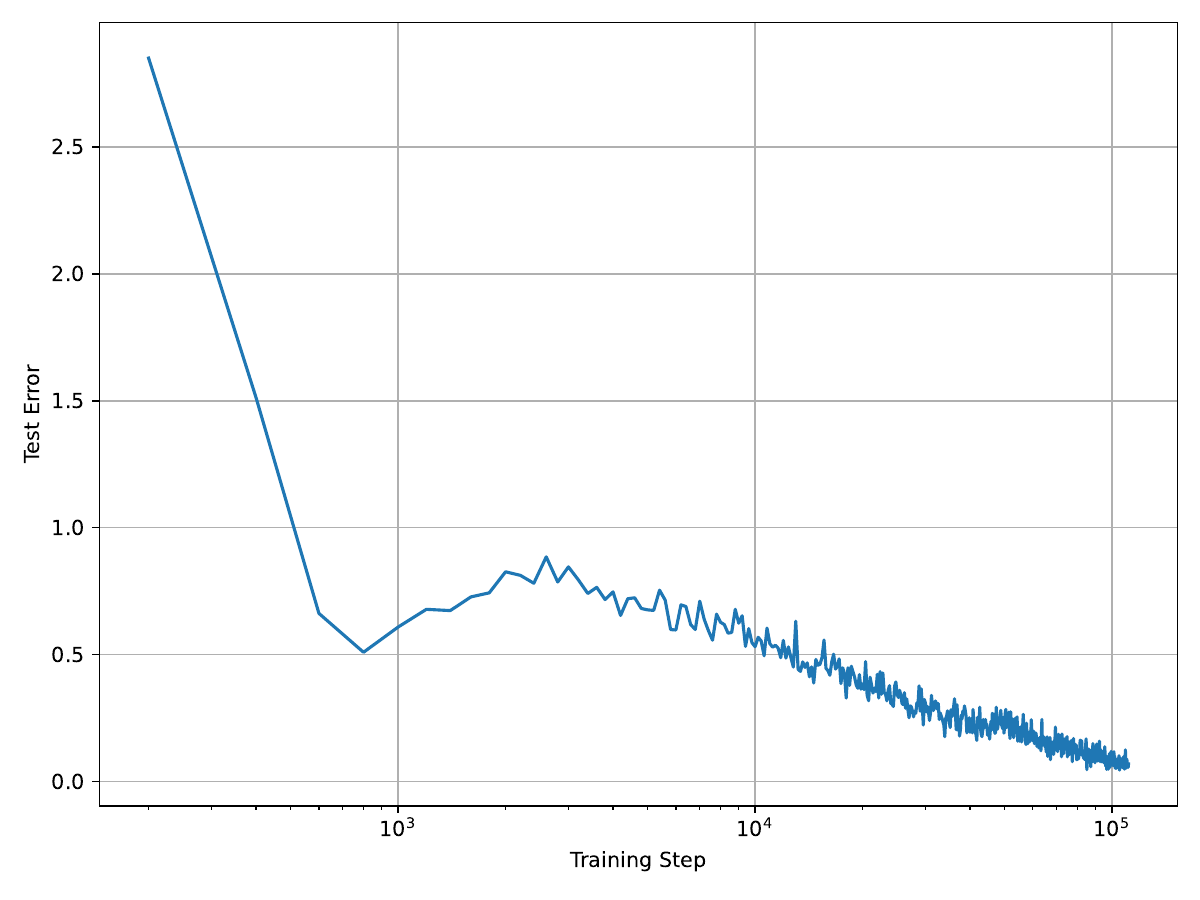}
        \caption*{\textbf{(c)}}
    \end{minipage}
    \caption{Results for division mod 96. In (a), we have plotted the training accuracy and the test accuracy as a function of the number of training steps which demonstrates grokking. In (b), we plot the corresponding trajectory on the information plane. In (c), we have plotted the test error as a function of the number of training steps which exhibits the double descent phenomenon. If we had plotted the figure using a linear scale, the separation between the two phases in (a) would be more pronounced. Further, note that the training error effectively vanishes at $\approx 5\times 10^3$ steps, but generalization only occurs at $\approx 10^5$ steps.} \label{Fig: moddiv Results}
\end{figure} 

\begin{figure}[h] 
    \centering
    \begin{minipage}[b]{0.4\textwidth}
        \centering
        \includegraphics[width=\linewidth]{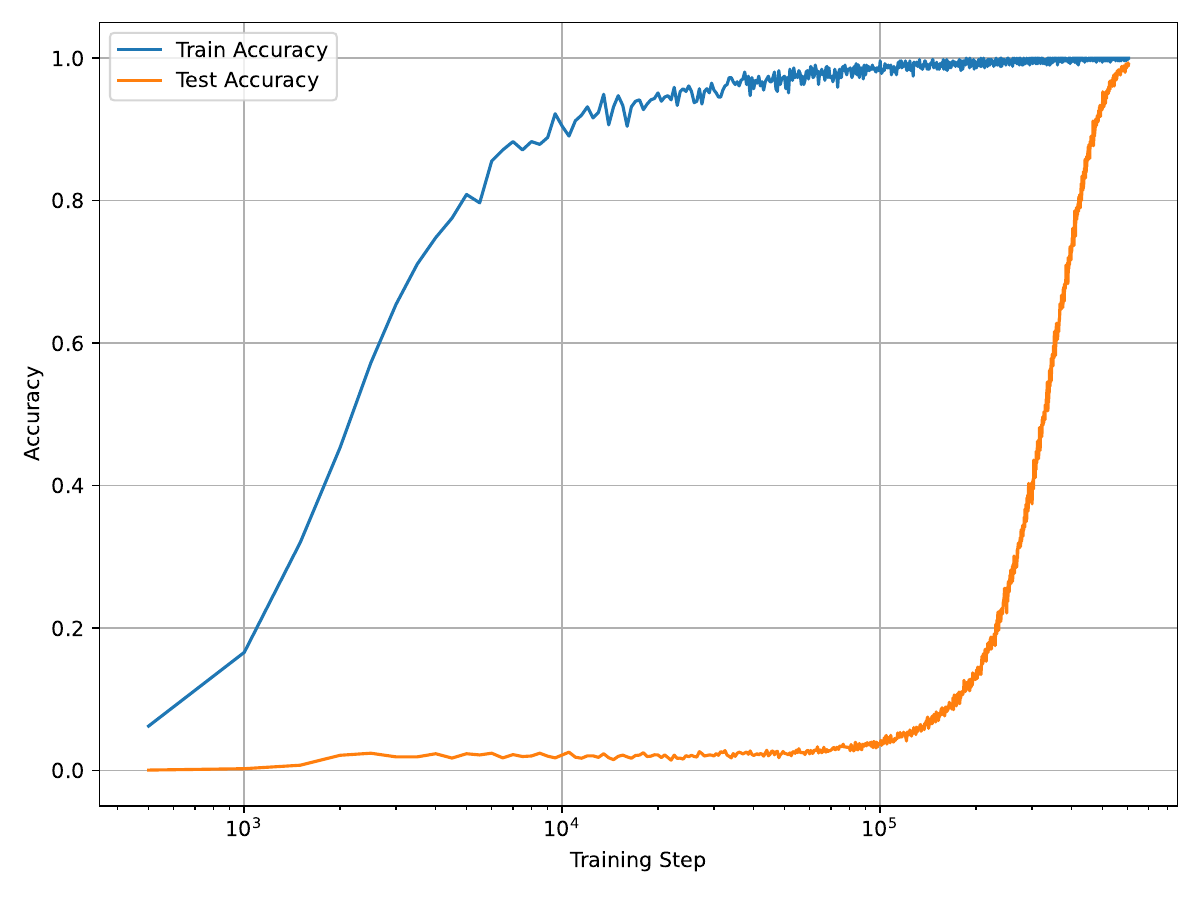}
        \caption*{\textbf{(a)} $S_5$ grokking}
    \end{minipage}
    \hfill
    \begin{minipage}[b]{0.4\textwidth}
        \centering
        \includegraphics[width=\linewidth]{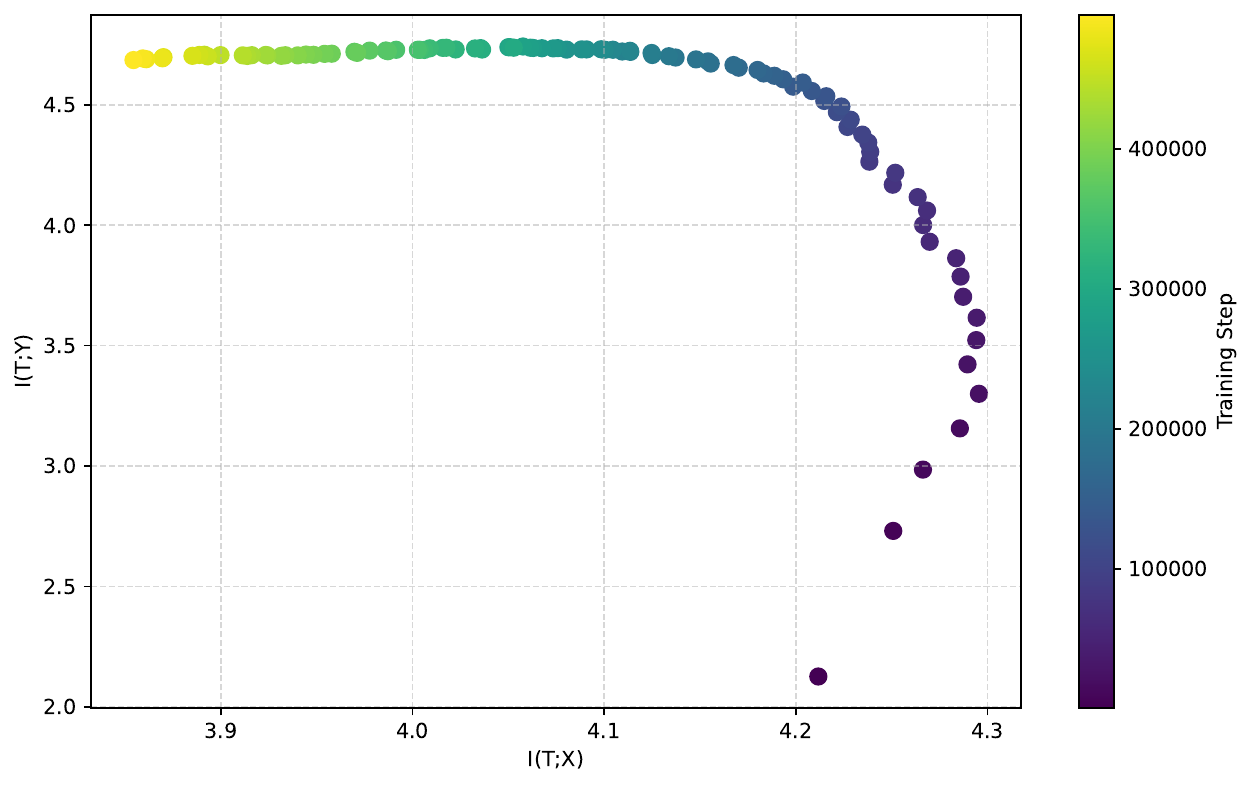}
        \caption*{\textbf{(b)} $S_5$ information bottleneck}
    \end{minipage}
    \hfill
    \begin{minipage}[b]{0.4\textwidth}
        \centering
        \includegraphics[width=\linewidth]{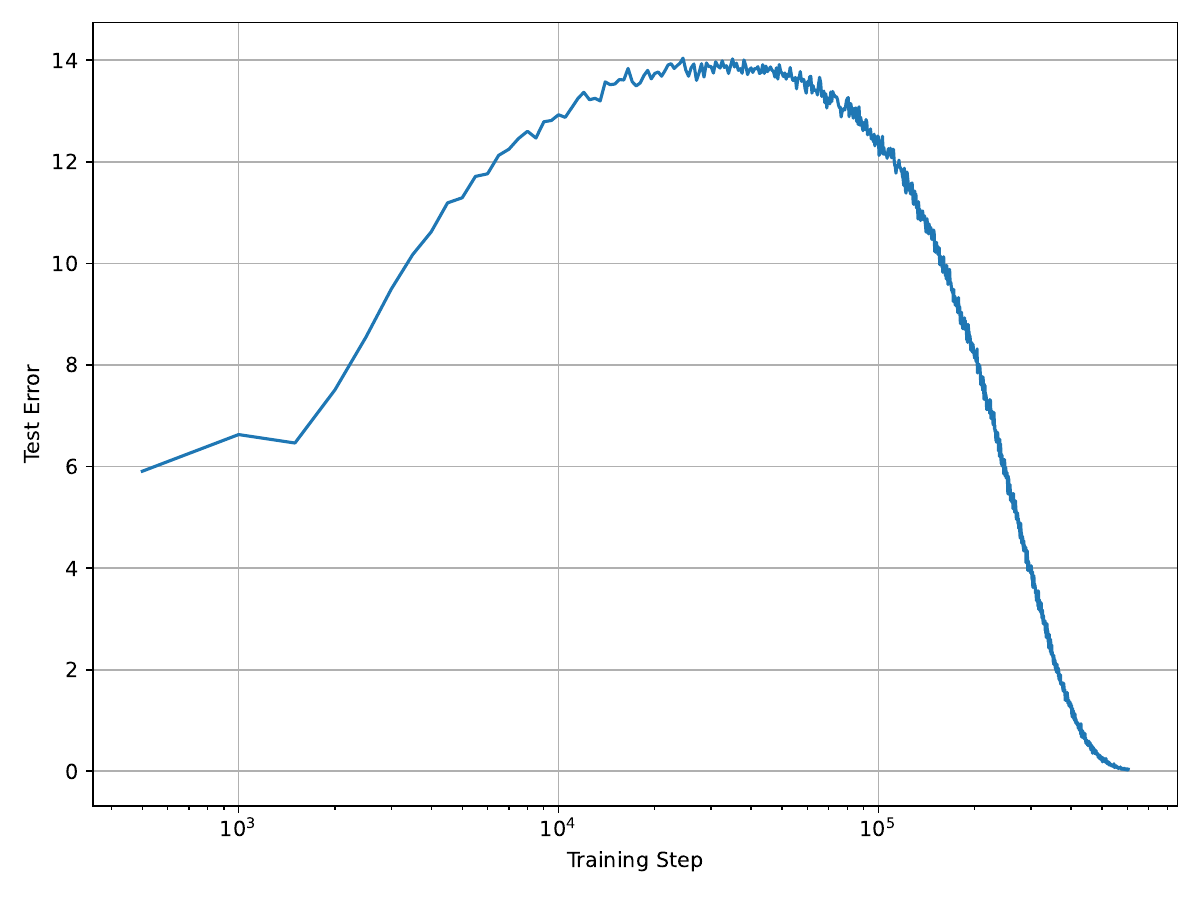}
        \caption*{\textbf{(c)} $S_5$ double descent}
    \end{minipage}

    \caption{Results for $S_5$. In (a), we have plotted the training accuracy and the test accuracy as a function of the number of training steps which demonstrates grokking. In (b), we plot the corresponding trajectory on the information plane. In (c), we have plotted the test error as a function of the number of training steps which exhibits the double descent phenomenon. In this case from figure (a) we see that the two phases of learning are well separated. This coincides with a pronounced hump before the onset of second descent in (c).} \label{Fig: S5 Results}
\end{figure}

\subsection{Results} \label{Results}
We present our results for the modular division dataset and $S_5$ datasets in Figures \ref{Fig: moddiv Results} and \ref{Fig: S5 Results} respectively. From Figure \ref{Fig: moddiv Results}, we see that midway between $10^3$ and $10^4$ steps, the fitting phase ends which is in good empirical agreement with when the model starts to generalize, marked both by the onset of the second dip of the double descent plot and the right-most point of the trajectory on the information plane. Also, we see that generalization ends at around $10^5$ steps, where the test accuracy reaches its maximum value. The point at which the test error vanishes corresponds to the point at which the mutual information trajectory hits a ``fixed point'' on the information plane. For the $S_5$ case (see Figure \ref{Fig: S5 Results}), grokking shows that generalization starts later at around $10^5$ steps which again coincides with both the rightmost point of the information plane trajectory of the information bottleneck picture and the onset of the second descent in double descent. The fixed point is reached past the midway point between $10^5$ and $10^6$ training steps in this case. The aforementioned agreement of the time scales for both datasets we have considered provides strong numerical evidence that the phenomena of grokking, information bottleneck and double descent are closely related to each other. It is interesting to note that the compression phase is much longer for the $S_5$ case than the modular division case. This is very likely because ``learning" the algorithm for the $S_5$ group is more difficult than doing the same for modular division.

\begin{figure}[h] 
    \centering

    \begin{minipage}[b]{0.4\textwidth}
        \centering
        \includegraphics[width=\linewidth]{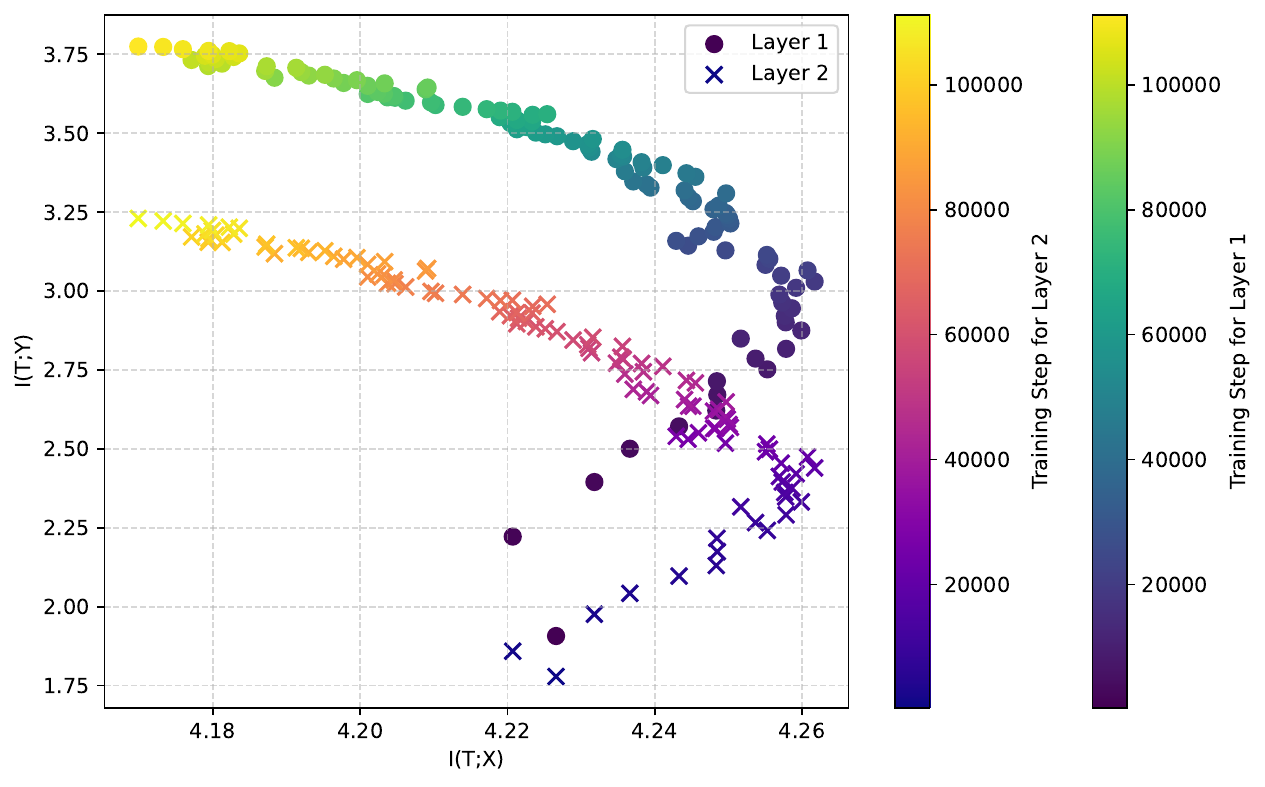}
        \caption*{\textbf{(a)} Evolution of mutual information for modular division}
    \end{minipage}
    \hfill
    \begin{minipage}[b]{0.4\textwidth}
        \centering
        \includegraphics[width=\linewidth]{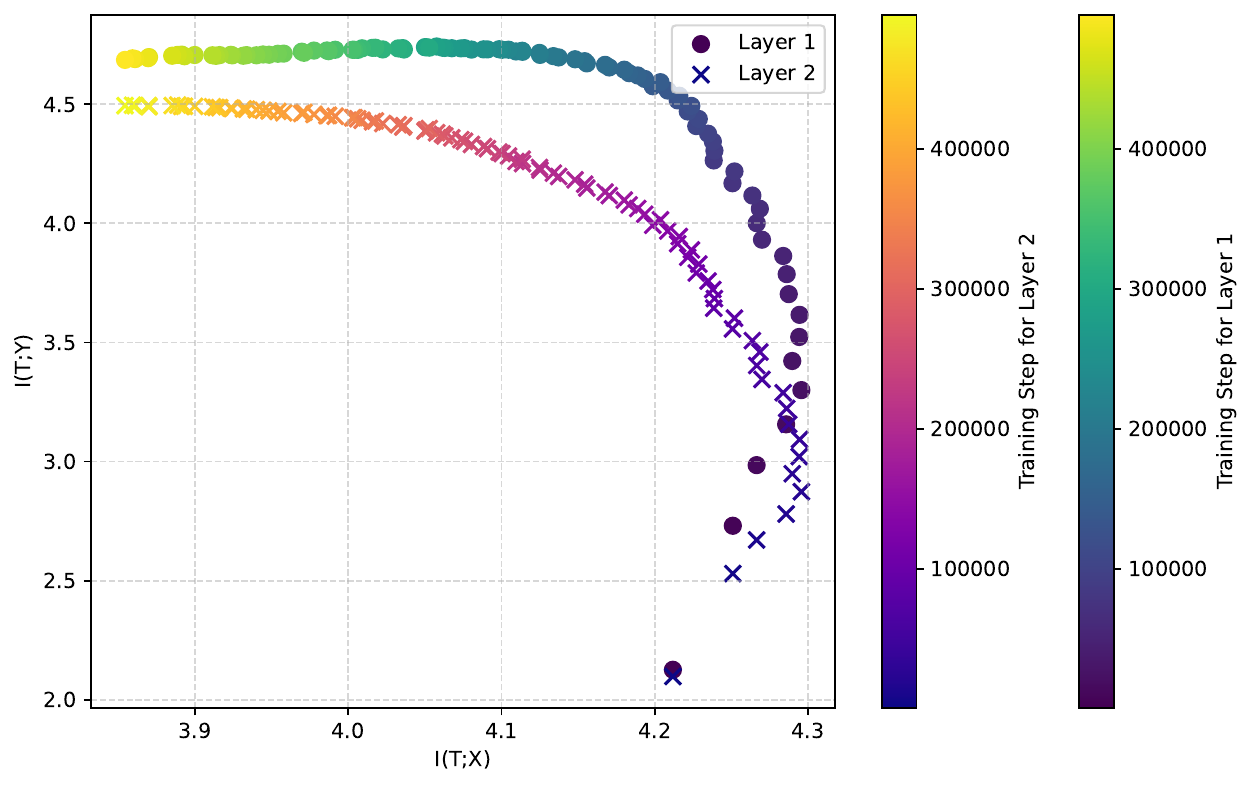}
        \caption*{\textbf{(b)} Evolution of mutual information for $S_5$}
    \end{minipage}

    \caption{Results for evolution of mutual information on the information plane for both modular division and $S_5$ datasets. Layer 1 lies in the first transformer block while Layer 2 in the second one. The onset of the forgetting phase is insensitive to the depth of the layer for both cases.} \label{Fig: MI comparison}
\end{figure} 

We have also explored if the value of the training step where the fitting phase ends is sensitive to the layer depth and the answer seems to be in the negative (see Figure \ref{Fig: MI comparison}).

\section{Two Phase Learning} \label{2-phase learning}

\subsection{Learning in two phases}

We have presented compelling numerical evidence for a meaningful connection between grokking, double descent, and the information bottleneck. In particular, we argue that these frameworks describe the same underlying two-phase learning process. During the first phase, the training loss is non-zero, and the network is actively driven to minimize it. This corresponds to the ascending limb of the classical U-shaped generalization curve described by double descent and reflects the standard bias–variance trade-off associated with overfitting during curve fitting. From the perspective of the information bottleneck, this phase is marked by an increase in the mutual information $I(T; Y)$, as the network accumulates knowledge useful for predicting the targets.

Once the grokking time scale is reached and the training error drops to zero, the loss function has been minimized. One might naively expect that learning ceases beyond this point. However, empirical observations show that the network continues to evolve in crucial ways during this phase, ultimately achieving generalization. Indeed, the second descent of the test error curve in double descent takes place well after the training loss has saturated. The information bottleneck framework offers a natural interpretation of this continued evolution: during this phase, the mutual information $I(T; X)$ is gradually minimized. This reflects a compression of the internal representations, whereby the network discards information about the input that is irrelevant to the task -- effectively ``forgetting'' what it does not need in order to generalize.

{\vskip 0.1cm}

Taken together, these insights motivate the following hypothesis about learning dynamics in deep neural networks: \emph{Learning in a DNN proceeds through two distinct phases — an initial phase of curve fitting, followed by a slower phase of ``coarse graining'' or compression. While the first phase is rapid and often occupies only a small fraction of total training time, the second phase is essential for generalization.}

\subsection{Connection to grokking dynamics}

This hypothesis is consistent with the dynamics of grokking described in Subsection~\ref{Grokking}. At the conclusion of the first phase of learning, the network is approaching the peak of the classical U-shaped generalization curve, signaling the onset of overfitting. From the observed grokking behavior, we know that the local complexity (LC) is highly concentrated around the training samples. This is exactly what one would expect from a network that has overfit the data: small perturbations of the training inputs can move the input across boundaries between distinct linear regions. As each region corresponds to a different effective model, the network’s predictions become highly sensitive to minor input changes—an archetypal symptom of overfitting. In this regime, generalization fails because the model is overly attuned to irrelevant fluctuations in the training data.

During the second phase of learning, the network undergoes a qualitative shift. The linear regions reorganize and expand, leaving larger, more stable zones around the training points. As a result, small perturbations to training inputs no longer cause transitions between different linear regimes. This enhanced stability leads to increased robustness and generalization. In the information bottleneck framework, this is reflected in a reduction of the mutual information $I(T; X)$, indicating that the network has begun to ignore unnecessary details about the input. The model is no longer encoding fine-grained idiosyncrasies of the training data, but instead compresses its representation to what is essential for prediction.

An important consequence of this analysis is that the mutual information $I(T; X)$ itself serves as a \emph{process measure} of learning. This perspective provides an information-theoretic interpretation of the second phase: it is a phase of forgetting, in which the network sheds irrelevant information inherited from the training data. The decrease in mutual information thus signals that the model is progressing toward effective generalization.

\subsection{Renormalization Group}

There is a well-established setting in theoretical physics where a process closely resembling the second phase of learning plays a fundamental role: the extraction of low-energy, long-distance effective dynamics from a detailed microscopic theory. This procedure is formalized through the \emph{renormalization group} (RG)~\cite{RG1}. A classic example involves a position-space lattice with a spin variable defined at each lattice site. One then performs a \emph{block-spin transformation}, in which nearby spins are averaged to produce a single effective spin representing a coarse-grained degree of freedom. Repeating this transformation iteratively leads to a description of the system at increasingly larger length scales, ultimately yielding the long-distance effective theory.

Several key features of the block-spin transformation are worth noting. First, it constitutes a literal \emph{coarse graining} of position space, whereby many microscopic degrees of freedom are replaced by a smaller number of effective variables. During this process, most of the parameters in the original microscopic theory shrink toward zero—these are the so-called \emph{irrelevant parameters}. Their decay is the mechanism by which the coarse-grained theory ``forgets" fine-grained details of the underlying system. A much smaller subset of parameters remain largely unchanged; these are \emph{marginal parameters}. Finally, an equally small subset of \emph{relevant parameters} are those whose influence grows under coarse graining. The marginal and relevant parameters control the large-scale behavior of the system.

The fact that the vast majority of microscopic parameters are eliminated under RG flow has profound implications: models with dramatically different microscopic structures can exhibit the same large-scale behavior. This is the physical basis of \emph{universality} in critical phenomena.

This analogy raises a natural question~\cite{RG2,RG2a,RG3,RG4,RG5}: could the renormalization group framework play an active role in the second phase of deep learning? During this phase, both the LMN and the network’s complexity decrease. This reduction in complexity constitutes a kind of compression: by the end of the second phase, the partition of input space consists of significantly fewer linear regions. Consequently, the number of distinct circuits used to construct the network’s function is also reduced. Fewer circuits imply fewer linear maps, which in turn suggests a reduced number of effective parameters -- mirroring the behavior of coarse graining in physical systems.

\subsection{Towards a theory of coarse graining}

In the second phase of learning, the network performs a form of coarse graining by discarding irrelevant details about the training data. This process is marked by a reduction in both the mutual information $I(T; X)$ and the number of linear regions partitioning the input space. In this section, we argue that this coarse-graining evolution proceeds via three fundamental mechanisms. Our goal is to lay the groundwork for an analog of the block-spin transformation in the renormalization group framework. The mechanisms we identify are naturally described within the circuit-based representation of DNNs and align with the observed decrease of associated progress measures such as local complexity and the linear mapping number.

\begin{figure*}[h]
\includegraphics[width=1.0\linewidth]{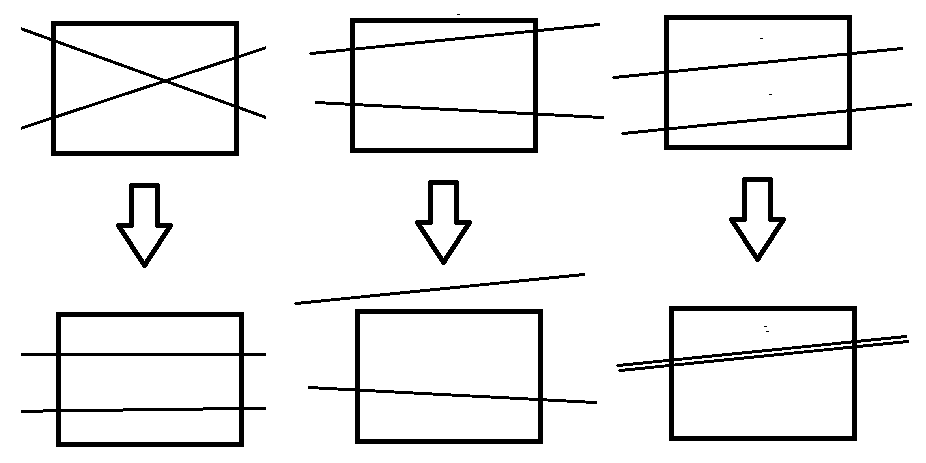}
\caption{\textbf{Evolution of neuron zero-locus lines:} Three ways in which the linear region count can decrease. \textbf{Left:} Two lines align, eliminating their intersection and reducing four regions to three. \textbf{Middle:} A line migrates out of the input space, collapsing three regions into two. \textbf{Right:} Two lines merge to form a single composite boundary, again reducing the region count from three to two.}
\label{fig:coarsegrain}
\end{figure*}

As the network trains, its weights and biases evolve continuously, inducing corresponding shifts in the preactivation zero-locus curves defined by Equation~(\ref{precativationlocus}). These zero-locus lines—which determine neuron activation boundaries—partition the input space into distinct linear regions, each associated with a specific circuit. During the second phase, the evolution of these lines must reduce the total number of linear regions, thereby driving the integrated local complexity toward zero. We identify three distinct geometric transformations by which this reduction can occur:
\begin{itemize}
\item \textbf{Alignment:} Zero-locus lines reorient so that they no longer intersect, eliminating vertices and reducing region count.
\item \textbf{Disappearance:} A line exits the input space entirely, effectively forcing the associated neuron into a single (active or inactive) state.
\item \textbf{Merging:} Two or more lines become coincident, creating composite decision boundaries shared by multiple neurons.
\end{itemize}
These processes are illustrated in Figure~\ref{fig:coarsegrain}.

When zero-locus lines align, their intersection point exits the input space, collapsing adjacent regions. Since each region corresponds to a unique circuit, the disappearance of a region implies that the network has discarded a previously active circuit.

In the case of disappearance, the zero-locus line simply moves outside the input domain. The corresponding neuron becomes either always active or always inactive throughout the relevant input space, and thus ceases to contribute meaningfully to computation. This reflects a form of neuron-level pruning emerging dynamically during training.

Finally, when two or more lines merge, they form a shared composite boundary. This boundary no longer corresponds to a single neuron’s activation threshold but instead to a coincident transition shared by multiple neurons. As a result, regions separated by such boundaries differ by multiple neurons. The compression process can thus yield a small number of surviving linear regions, each defined by qualitatively distinct circuits.

The Figure 2 of \cite{Grok5} shows a layerwise visualization of the input space formation of circuits on a two dimensional subspace, which passes through three training samples, for a network after grokking. Layer 5 in particular shows the formation of complex composite boundaries which localize around the decision boundary.

The discussion of this section suggests that the circuit framework provides a useful basis for developing a dynamical theory of the second phase of learning.

\section{Discussion} \label{Discussion}

In this article, we have proposed that learning in deep neural networks proceeds through a two-stage process: an initial phase of curve fitting, followed by a slower phase of coarse graining or compression. The first phase is typically rapid -- often completed in a small fraction of the total training time -- as the network efficiently minimizes the loss function and achieves near-zero training error. The second phase unfolds more gradually: it is during this period that the network refines its internal representations, discards irrelevant input details, and develops the ability to generalize.

An important result of our analysis is that the mutual information $I(T; X)$ serves as a \emph{process measure} of learning: decreasing mutual information signals that the model is progressing toward generalization.  

It is plausible that the second phase is prolonged simply because it emerges as a side effect, rather than an explicit objective of the training procedure. Standard training algorithms are designed to minimize the loss function and maximize test accuracy, effectively optimizing for curve fitting. In contrast, the compression phase -- during which the network performs a kind of coarse graining -- is likely incidental. The training dynamics are not explicitly constructed to achieve this stage, yet it arises nonetheless. With a deeper understanding of the mechanisms that govern this second phase, it may be possible to develop tailored training algorithms that deliberately promote compression, thereby significantly reducing the duration of this phase and accelerating the path to generalization.

We have taken initial exploratory steps toward employing the circuit framework as a foundation for developing a dynamical theory of the second phase of learning—a phase characterized by the progressive reduction of network complexity. This framework naturally directs attention to the evolution of the boundary of linear regions, with each linear region associated with a distinct circuit. Within this perspective, we have identified and explained three key mechanisms that contribute to complexity reduction: (1) the elimination of specific circuits as linear regions vanish, (2) the effective removal of neurons whose activation boundaries exit the input space, and (3) the merging of decision boundaries that allows structurally distinct circuits to become adjacent through the formation of composite interfaces. These insights suggest that the circuit-based view offers a principled way to understand the structural reorganization that occurs during compression. Further developing this perspective and applying it to detailed studies of the second phase dynamics presents a promising direction for future work.

\begin{center} 
{\bf Acknowledgements}
\end{center}
This work is supported by a start up research fund of Huzhou University, a Zhejiang Province talent award and by a Changjiang Scholar award.

\end{document}